\documentclass[journal,draftcls,onecolumn,11pt]{IEEEtran} 
\usepackage{cite}
\usepackage{graphicx}
\usepackage{psfrag}
\usepackage{subfigure}
\usepackage{url}
\usepackage{amsmath}
\usepackage[normalem]{ulem}
\usepackage{epsfig,colortbl}
\usepackage{amssymb,comment}
\usepackage{enumerate}
\usepackage{times}
\usepackage{multirow,multicol}
\usepackage[ruled, vlined]{algorithm2e}
\usepackage{tabularx}
\usepackage{todonotes}

\newcounter{ctr}

\makeatletter

\newcommand{\be}{\begin{eqnarray}}
\newcommand{\ee}{\end{eqnarray}}

\begin{document}
\title{ Device Fingerprinting in Wireless Networks: Challenges and Opportunities}
\author{Qiang Xu$^\dag$, Rong Zheng$^\dag$, Walid Saad$^\ddag$ and Zhu Han$^*$ \\
$^\dag$Department of Computing and Software, McMaster University, Hamilton, ON, Canada\\
Email:\{xuq22, rzheng\}@mcmaster.ca \\
$^\ddag$Bradley Department of Electrical and Computer Engineering, Virginia Tech, VA, USA \\
Email: walids@vt.edu \\
$^*$Department of Electrical Engineering, University of Houston, Houston, TX, USA \\
Email: zhan2@mail.uh.edu
}

\maketitle 
\begin{abstract} 
Node forgery or impersonation, in which legitimate cryptographic credentials are captured by an adversary, constitutes one major security threat facing wireless networks. The fact that mobile devices are prone to be compromised and reverse engineered significantly increases the risk of such attacks in which adversaries can obtain secret keys on trusted nodes and impersonate the legitimate node. One promising approach toward thwarting these attacks is through the extraction of unique fingerprints that can provide a reliable and robust means for device identification. These fingerprints can be extracted from transmitted signal by analyzing information across the protocol stack. In this paper, the first unified and comprehensive tutorial in the area of wireless device fingerprinting for security applications is presented. In particular, we aim to provide a detailed treatment on developing novel wireless security solutions using device fingerprinting techniques. The objectives are three-fold: (i) to introduce a comprehensive taxonomy of wireless features that can be used in fingerprinting, (ii) to provide a systematic review on fingerprint algorithms including both white-list based and unsupervised learning approaches, and (iii) to identify key open research problems in the area of device fingerprinting and feature extraction, as applied to wireless security. 
\end{abstract}


\section{Introduction} 
\label{sect:intro} 
With the proliferation of mobile devices and the advent of the Internet of Things, wireless technologies are becoming essential parts in modern computing platforms and embedded systems to provide low-cost, anytime, and anywhere connectivity. While wireless networks share many of the same vulnerabilities as wired networks and can be often used as a stepping stone for attacks at a larger scale, the broadcast nature of the wireless transmission medium tends to aggravate the problems by making it easier to compromise the service accessibility as well  confidentiality and integrity of data communication.

Many cryptography-based approaches~\cite{sklavos2003mobile} exist for authentication, data confidentiality, and integrity in wireless networks. However, such techniques are powerless in face of denial-of-service attacks (DoS) such as jamming. Furthermore, implementations of wireless security protocols are known to be plagued with security holes that can be easily exploited. Examples are statistical analysis in Wired Equivalent Privacy (WEP)~\cite{wep}, recovery of passphrases in Wi-Fi Protected Access (WPA) and Wi-Fi Protected Access II (WPA2)~\cite{mavridis2011real}, man-in-the-middle attacks in cellular networks~\cite{meyer2004man}, among others. Finally, many wireless networks such as wireless mesh networks, wireless sensor networks, cognitive radio networks, small cell networks assume a degree of cooperation among users. This makes them particularly vulnerable to forgery and insider attacks once the malicious users obtain security credentials from legitimate users and become part of the networks.  The fact that mobile devices are prone to hacking, compromising, and reverse engineering, coupled with poor security management mechanisms in wireless network systems, significantly increases the risk of attacks.  Therefore, novel and low-complexity methods for efficiently identifying legitimate users to detect potential attacks from malicious adversaries are of great importance.

Recently, device fingerprinting, the process of gathering device information to generate device-specific signatures and using them to identify individual devices, has emerged as a promising solution to reducing the vulnerability of wireless networks to node forgery or insider attacks\cite{Bratus2008,Brik2008,Chen,Desmond2008,Franklin,Gao,Hall2005,Nguyen2011,Scanlon2010,Ureten2007,Neumann}. The basic idea is to passively or actively extract unique patterns (also called {\it features}) manifested during the process of wireless communication from target devices. A variety of features can be extracted and utilized including physical layer (PHY) features, medium access control (MAC) layer features, and upper layer features.  Effective device fingerprints must satisfy two properties that include: i) they are difficult or impossible to forge, and ii) the features should be stable in the presence of environment changes and node mobility. The first requirement renders identifiers such as IP addresses, MAC addresses, electronic serial number (ESN), international mobile station equipment identity (IMEI) number, or mobile identification number (MIN). Unsuitable candidates as all these identifiers have been shown to be easily modifiable via software~\cite{douceur2002sybil,yang2008detecting}. In contrast, location-dependent features such as the popular radio signal strength (RSS) cannot be used on their own as fingerprints as they are susceptible to mobility and environmental changes.

Despite many interests in wireless device fingerprinting and their potential in enhancing wireless security, surprisingly, the existing literature is sparse with no comprehensive overview on the state of the art and the key fundamentals involved. The main \emph{contribution} of this paper is thus to provide a detailed survey of features and techniques that can be adopted in wireless device fingerprinting. We introduce a comprehensive taxonomy of wireless features that can be used in fingerprinting.Specifically, we classify the features based on the layer of protocol stacks they are generated from, whether they are active or passive, and the granularity they can work at. In addition, the fingerprinting algorithms are also systematically reviewed.

The rest of the paper are organized as follows. In Section~\ref{sect:overview}, we first motivate the needs for device fingerprinting and then give an overview of the basic procedure. In Section~\ref{sect:feature}, a taxonomy of features that can be utilized in fingerprinting approaches is provided.  The advantages and disadvantages of different features are also discussed. In Section~\ref{sect:algo}, security algorithms for device identification are discussed in detail. In particular, we break down existing fingerprint algorithms into white-list based and unsupervised learning based approaches. In Section~\ref{sect:future}, potential research directions are discussed, and finally, we conclude the paper in Section~\ref{sect:conclusion}.

\section{Overview of Wireless Device Fingerprinting}
\label{sect:overview}
\subsection{Motivation}
\label{sect:need}

Node forgery or identity spoofing in itself may not cause significant harm to the operations of wireless networks. However, combined with other attack tactics, node forgery can be used to launch more sophisticated attacks that may greatly compromise the networks' serviceability and confidentiality. For example, in the authentication and association flooding attack in WLANs~\cite{liu2006analysis}, a malicious device sends out authentication or association frames at short intervals to overload the access point (AP) or the authentication server.  If the malicious device uses a single MAC address, the attack can be mitigated by blocking the MAC address (though DoS may still occur due to jamming). However, it is much harder to detect and block if the attacker changes its MAC address in each request, mimicking the existence of many clients.

To compromise the confidentiality of data, a malicious user can set up a rogue base-station or AP (e.g., an AP  installed without the authorization of the system administrators). Instead of communicating with a legitimate AP, connecting to a rogue AP can result in the interception of data from clients.  Though the IEEE 802.11x has been deployed in conjunction with WPA or WPA2 to authenticate APs in WLANs, many enterprise networks use self-generated certificates (rather than those from a certified authority), and users do not always verify the legitimacy of the certificate provided. Similar problems exist in cellular systems~\cite{Barbeau}. Many tools are available such as Airsnarf~\cite{Airsnarf} and Raw-FakeAP~\cite{RawFake} that can be used to deploy rogue APs in WLANs.  Setting up multiple rogue APs by using multiple MAC addresses on a single physical device can increase the possibility of having legitimate users associated with one of the rogue APs, and thus causes greater harm to data confidentiality.

A naive solution to defending against node forgery is to verify the MAC address of the device against the legitimate ones. However, such ``software" based identifiers are very easy to spoof. For example, an \textit{ioctl} system call in UNIX-like operating systems can modify the MAC address of a network interface card. Modifying or replacing the Erasable Programmable Read Only Memory (EPROM) in a phone would allow the configuration of any ESN and MIN via software for cellular devices. The ease of compromising firmware or software identifications remains a serious threat to the security of most wireless systems.

\subsection{Enhancing wireless security with device fingerprints}

A device identification system consists of a profiler that extracts device fingerprints, targeted devices and in some cases, a database that stores the fingerprints associated with legitimate devices. Despite a variety of existing work on using fingerprints to enhance wireless security, they generally follow three common steps, namely, i) identifying relevant features, ii) extracting and modeling features, and iii) device identification.  In what follows, we elaborate more on each of these steps.

\paragraph*{Step 1: Identifying relevant features}

Signal transmission in wireless communication offers a rich set of features that can be utilized for the purpose of device identification. Relevant features can be found at all layers of the protocol stack. Among device-specific features, one can explore characteristics such as clock skew observed from the time stamps of messages in MAC layer frames, packet inter-arrival times in the MAC or upper layers, and various physical layer RF parameters (e.g., the transient phase at the onset of transmissions, frequency offsets, and phase offsets, etc.). Among location-dependent features, one can use RSS or channel state information (CSI) measured from trusted devices.  Furthermore, multiple features can be combined to form a device fingerprint.  Different features may have different granularities in device identification giving rise to trade-offs in false positive and false negative rates.  For instance, location-dependent features such as RSS must be used in conjunction with other credentials (e.g, device ID and MAC address) to provide effective identification.

This step is different with feature selection which itself is a challenging problem and is dependent on the hardware capability of the system -- an issue that will be discussed in Section~\ref{sect:feature}.

\paragraph*{Step 2: Extracting and modeling features}

In this step, the features are extracted from raw observations from the transmitted signal of targeted devices.  While some features such as RSS are readily available and can be easily measured, profiling more complex features such as frequency or phase shift differences requires more sophisticated signal processing techniques and/or specialized hardware.  Due to the dynamic nature of wireless communication channels and imperfection in the profiler implementation, the resulting features tend to be stochastic. A proper stochastic model needs to be constructed for device identification.

\paragraph*{Step 3: Device identification}

Once features are extracted and mapped to a stochastic model, the next step is to develop the machine learning algorithms that utilize these features for device identification. Depending whether prior information exists regarding the fingerprints of legitimate devices (called {\it white-list}), the algorithm can be either supervised or unsupervised. In supervised approaches, the newly obtained device fingerprint is compared against those in the white-list. If the new fingerprint deviates significantly from the known ones, an attack is detected. In unsupervised approaches, similar fingerprints from multiple {\it logical} devices (as indicated by their identifiers) are grouped together and mapped to the same physical device.

\begin{figure}[t!]
\centering
\includegraphics[width=100mm]{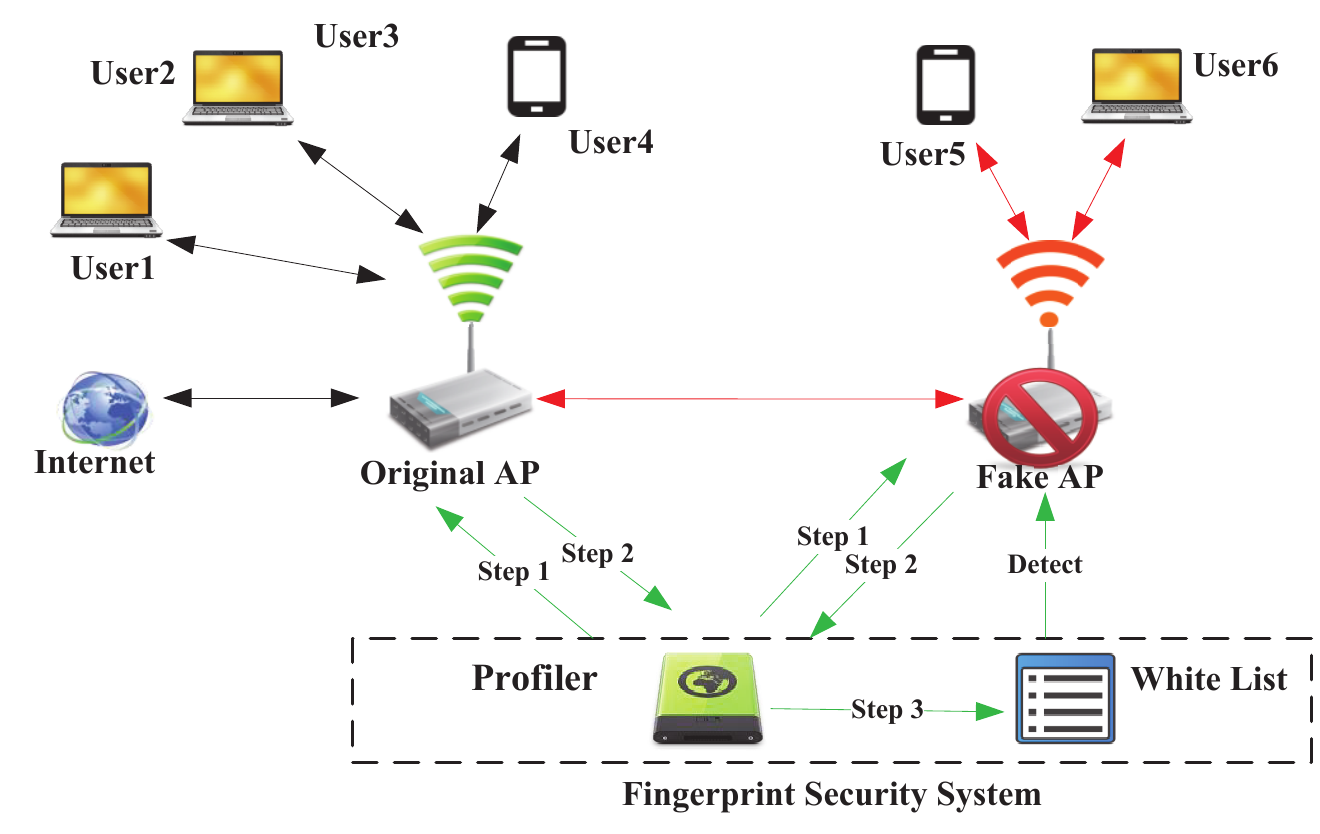}
\caption{The application of fingerprints in detecting rogue APs}
\label{fig:overview}
\end{figure}

As an example, consider the use of device fingerprinting to identify rogue APs.  In Figure~\ref{fig:overview}, a malicious attacker masquerades as an AP. User 5 and User 6 are associated with the rogue AP to access the Internet and their data confidentiality is compromised.  As shown in Figure~\ref{fig:overview}, by introducing the fingerprinting security system, the attacker can be detected.  The profiler first identifies (step 1) and extracts features (step 2) to generate the fingerprints of legitimate APs, which are stored in a white-list. To determine whether an AP is rogue or not, one can scan the white-list to see whether a similar fingerprint exists.

\section{A Taxonomy of Features for Device Fingerprinting}
\label{sect:feature}

A variety of features can be utilized for device fingerprinting. In this section, we provide a taxonomy of features proposed in literature.  In particular, we classify the features based on the layer of protocol stacks they are generated from, whether they are active or passive, and the granularity they can work at. In the subsequent discussion, we use the terms ``feature" and ``signature" interchangeably.

\subsection{Features in different layers}
\subsubsection{PHY layer}
PHY layer features are derived from the received RF waveform.  They generally fall into two categories: (i) Location dependent features; and (ii) Location independent features (also known as {\it radiometrics}).

\paragraph*{Location-dependent features}

RSS is the most commonly used location-dependent feature. It is typically reported as a single number (e.g., in dB and dBm) by  wireless device drivers.  RSS measures the average signal power at the receiver and depend on the transmission power at the sender and the attenuation in the channel. Two distant locations can have very different RSS values with respect to the same transmitter. On the other hand, if two devices are in close proximity, their RSS values tend to be similar. Another more fine-grained location-dependent feature is the channel state information at the receiver (CSIR). Due to small-scale fading, CSIR can differ a lot when a receiver moves by only a fraction of the wavelength. Location-dependent features alone are insufficient for fingerprinting as devices may move and the channel condition changes over time.

\paragraph*{Location-independent features}

Location-independent features pertain to the hardware implementation of individual devices (chipsets). Despite significant advancement in micro-electronic circuit design and manufacturing, there are still imperfections in the manufacture process of wireless transmitters\cite{Brik2008,Polak2011,Dolatshahi2010}.  Such imperfections result in broad variations in key device parameters (e.g.  channel width, channel doping, concentration, and oxide thickness) among production lots.  These variations, though small enough to meet the specifications of communication standards and certification requirements, allow for unique characterization of the devices and form device fingerprints.

Hall {\it et. al}~\cite{Hall2005} found that the unique characteristics of a transceiver are manifested in the turn-on transient portion of signals.  The difference in these transient behaviors among different devices is observable even for transmitters from the same manufacture lending them good candidates for fingerprinting. In \cite{Ureten2007,Hall2005}, once the transient phase is isolated, amplitude, phase angle and frequency are extracted as features using the Discrete Wavelet Transform (DWT), where amplitude and phase are calculated as $\alpha(t) = \sqrt{I^2(t) + Q^2(t)}$, $\theta(t) = tan^{-1}[\frac{Q(t)}{I(t)}]$ and $I(t)$ and $Q(t)$ are the in-phase and quadrature components in the signal over time.

In \cite{Polak2011,Dolatshahi2010}, Polak, Dolatshahi and Goeckel exploited power amplifier imperfections to identify wireless devices. Since power amplifiers are the last elements in the RF chain of transmitters, it is especially hard for attackers to modify via software. Specifically, the nonlinear characteristics of power amplifiers are modeled with Volterra series representations~\cite{Polak2011,Dolatshahi2010}. The Volterra coefficients capture the I/O characteristic of the associated amplifier, and form device fingerprints.

Brik {\it et al.} \cite{Brik2008} implemented a system named ``PARADIS'' that made use of features such as magnitude and phase errors, I/Q origin offset and SYNC correlation of the frame in question. In \cite{Nguyen2011}, Nguyen {\it et al.}  exploited carrier frequency difference (CFD) $\delta f_c$ and phase shift difference (PSD) $\phi$ to fingerprint wireless devices.  $\delta f_c$ is the difference between the carrier frequency of the ideal signal and the one of the transmitted signal which is likely to be different for different wireless transmitters. PSD $\phi$ is defined as the phase shift from one constellation to a neighboring one which may vary because of the difference between the transmitter amplifier for I-phase and Q-phase. Nguyen {\it et. al}~\cite{Nguyen2011} further considered the use of second-order cyclostationary feature (SOCF) in conjunction with PSD and CFD to identify devices that employ OFDM transmission technologies.

In extracting RF features, radiometric techniques can be classified into waveform domain and modulation domain based on how they treat signals\cite{Brik2008}. Waveform domain techniques~\cite{Hall2005,Remley,Sheng,Ureten2007,Polak2011} consider time and frequency representation as the basic blocks while modulation domain techniques~\cite{Brik2008} represent signals  in terms of I/Q samples. Waveform domain techniques are more flexible but more complex. Modulation domain techniques are better structured and well-behaved but require knowledge of the respective modulation scheme.

\subsubsection{MAC layer}

Exploring features at MAC layer has also attracted a lot of attention because such features are relatively easy to extract and do not require specialized hardware. In general, the key idea behind fingerprinting using MAC layer characteristics is that some details are underspecified in the standards, and the implementation of these details was left to the vendors. Thus, MAC layer features are usually vendor specific.

In \cite{Desmond2008}, Desmond {\it et al.} characterized the active scanning process in the IEEE 802.11 networks as cycles, where each cycle consists of: i) a rapid burst of zero or more probe requests with small time intervals in the range of milliseconds; and ii) a probe request after a prolonged time period in the range of tens of seconds. The large delays between cycles are termed as ``inter-burst latencies'' and can be used for identification. Franklin {\it et al.} \cite{Franklin} derived features based on the bin frequency of arrival time between probe request frames during active scanning. Corbett {\it et al.} also observed unique traffic patterns due to differences in the implementation of the active scanning function~\cite{Corbett2006,Corbett2008a}.  They applied spectral analysis and used frequency-domain features to detect unauthorized users.

To effectively detect MAC address spoofing, Jana and Kasera~\cite{Jana} calculated the clock skew of an AP from the IEEE 802.11 Time Synchronization Function (TSF) timestamps sent in the beacon/probe response frames and use it as device feature. Arackaparambil {\it et al.}\cite{Arackaparambil2006} analyzed the robustness of such an approach and identified some deficiency.  In particular, they found that upon synchronization with a legitimate AP, a device can ``acquire" the same clock skew as the AP. Therefore, it is not possible to detect a fake AP by comparing the clock skew alone, with a reasonable degree of certainty.  Instead, the authors suggest to use the line-fitting error and jitter of the beacon timestamps as features.

In \cite{Corbett2008}, Corbett {\it et al.} explored the underspecified rate switching mechanism in the IEEE 802.11 standard. They recorded the data rate information contained in the 802.11 PHY layer header, and analyzed the occurrence of rate switching events using spectral analysis. The periodicity embedded in the wireless traffic caused by rate switching is used as device fingerprint.

Cache~\cite{Cache} proposed two methods for fingerprinting a device’s network card and driver. The first one uses the 802.11 association redirection mechanism.  Though well specified in the IEEE 802.11 standard, it is very loosely implemented in the tested wireless cards. As a consequence, each wireless card behaves differently during this phase which can set them apart.  The method analyses the duration field values in 802.11 data and management frames.  Wireless cards from different vendors compute the duration field in slightly different ways.

Another source of MAC layer features can be derived from the responses of wireless interfaces to non-standard events. Bratus {\it et al.}\cite{Bratus2008} analyzed the IEEE 802.11 MAC header fields, and singled out combinations of frame type, subtype and Frame Control flags that are either prohibited by the standard, not explicitly prohibited but are of unclear utility, or simply unlikely to occur in practice. They built BAFFLE, a frame generator and injector that craft non-standard and malformed frames and elicit the responses from targeted devices. The responses then form features for device fingerprinting.

Neumann {\it et al.}~\cite{Neumann} evaluated the effectiveness of various MAC layer features for 802.11 devices including, transmission rate, frame size, medium access time, medium access time (e.g., backoff mechanism), transmission time, and frame inter-arrival time. Two criterion were considered, i) similarity of signatures from the same device generated at different time, and ii) dissimilarity of the signatures from different devices. The authors found that the network parameters such as transmission time and frame inter-arrival time perform the best in comparison to the other network parameters considered. The authors suggested the combination of multiple parameters for device fingerprinting as a future research direction.

\subsubsection{Network and upper layer features}
In contrast to abundant PHY and MAC layer features, features in other
layers are quite limited.

In \cite{Gao}, Gao {\it et al.} use TCP or UDP packet inter-arrival time (ITA) from APs as signatures to distinguish AP types. In particular, for each AP, multiple packet traces are collected. After ITAs are computed, the time series are sampled using bin sizes from $1\mu s$ to $10\mu s$. The optimal bin size is chosen that maximizes the difference in resulting ITAs among different APs. Strictly speaking, ITA is not a transport layer feature as it is useful only when packets from upper layers are tightly clustered (also called ``packet train") and the spacing in time for transmissions over the air is determined by the backoff timer implementation of APs.

In~\cite{M.EjazAhmed2014}, the traffic patterns are used as the features. The cognitive radio users perform clustering over the primary user traffic patterns, so as to optimize their transmission strategies. Similar idea can also be utilized for security purposes. For example, in digital TV broadcasting, the traffic patterns are very unique. If an attacker with very different traffic patterns is detected, security alarms can be raised.

Another high-level feature that may be exploited in device identification is browser signatures. Eckersley~\cite{Eckersley} shows that modern web browsers provide the version and configuration information  to websites upon request. These information can be used to track individual browser. It has been found that only two in 286,777 browsers may share a common signature. Since multiple browsers may run on the same machines and browser signatures can be collected via wired networks as well, they are out of the scope of this paper. Moreover, some literatures use behavioral patterns of users as fingerprints~\cite{Hsu2007,papadopouli2005characterizing}. For example, in~\cite{Hsu2007}, Hsu {\it et al.} analyze wireless users' association patterns by mining wireless network log from two major university campuses. They  find qualitative commonalities of user behaviors from the two universities. Papadopouli {\it et al.} characterize and analyze usage patterns of mobile users on campus WiFi network~\cite{papadopouli2005characterizing}. They found that session and visit durations are affected by mobility patterns and building types. These patterns may be used to distinguish devices. However, to our best knowledge, such application layer features have not been explored in wireless security context.

\subsection{Vendor specific vs. device specific features}

As previously discussed, some fingerprinting features are vendor specific and may change depending on the model and firmware version of the devices; while other features are device specific and are likely to differ even among devices from the same vendor.

PHY layer features utilize imperfection in the manufacturing process of individual chipsets, and are thus device specific. However, MAC layer features mainly take advantage of underspecified aspects of wireless standards as discussed earlier. Therefore, most MAC layer features are vendor specific, with the exception of clock skews in \cite{Brik2008,Dolatshahi2010,Polak2011}, which is device specific if the timestamps are generated from a local oscillator.

Clearly, vendor specific features are of higher granularity in device identification. However, since most of them can be extracted by crafting or inspecting MAC layer frames, commercial-off-the-shelf hardware would suffice in the implementation of such profilers.

\subsection{Features extracted via passive or active methods}

Another way to classify features for device fingerprinting is based on the extraction methods. There are two categories, namely, passive and active approaches. In passive approaches, the profiler observes the ongoing communication of the targeted system and extracts the needed features from the transmitted signal or packet/frame traces. In contrast, active approaches, inject signals or probe messages to elicit responses from devices to obtain useful features. With a slight abuse of terminologies, we call features extracted via passive and active methods, passive and active features, respectively.

\subsubsection{Passive features}

Among the features discussed thus far, rate switching, active scanning, the clock skew from time synchronization function (TSF) stamps, various radiometric features, random back-off times, duration field values in 802.11 data and management frames, ITAs are all passive features.

Since passive approaches do not inject any stimulant into the system, the extraction of passive features will not alert or disturb the system being surveilled. The benefits are two-fold. First, no additional medium contention or network congestion will be introduced. Second, the attackers would not be able to detect the existence of the defense mechanism.

\subsubsection{Active features}

Active approaches involve interrogating a node with various types of packets. These packets may vary in size, and can be either legitimate or malformed. The work by Bratus {\it et al.} that triggers responses from WiFi devices using crafted frames with rare combinations of header fields is one such approach. The feature extracted by exploiting 802.11 association redirection mechanism~\cite{Cache} requires the transmissions of association responses with different source addresses, and is thus of active nature. Compared to passive approaches, active feature extraction is less covert.  However, it allows the extraction of features that are not otherwise feasible.

\subsection{Summary}

In this section, we have provided a taxonomy of features that can be used for device fingerprinting. The features can be categorized based on which layer of the network stack they are from, whether passive or active extraction methods are applied, and whether they are targeted for device or vendor identification.  A comparison of various features is provided in Table~\ref{big table}, detailing which categories they belong to.

\begin{table}
\centering
\caption{The distribution of some selected features in different dimensions}
\label{big table}
\begin{tabularx}{\textwidth}{|X|c|c|c|}
\hline
\textbf{Features} & \textbf{Active or Passive} & \textbf{Layer}  & \textbf{Granularity} \\
\hline
Packet inter-arrival time~\cite{Gao} & Passive & MAC/Transport layer & Type of AP \\
\hline
Responses to crafted non-standard or malformed 802.11 frames\cite{Bratus2008} & Active & MAC & Type of wireless device \\
\hline
Implementation of rate switching\cite{Corbett2006a,Corbett2008} & Passive & MAC & Type of NIC \\
\hline
Implementation of active scanning\cite{Corbett2006,Desmond2008,Franklin,Corbett2008a} & Passive & MAC & Type of NIC \\
\hline
The clock skew of an AP from time synchronization function (TSF) stamps \cite{Jana,Arackaparambil2006} & Passive & MAC & Individual AP \\
\hline
Combined features\cite{Chen,Scanlon2010} & Passive & --- & Individual device \\
\hline
Various radio frequency features\cite{Remley,Brik2008,Sheng,Ureten2007,Dolatshahi2010,Hall2005,Polak2011} & Passive & PHY & Individual device \\
\hline
Random back-off timers\cite{Neumann} & Passive & MAC & Type of device \\
\hline
Implementation of association redirection mechanism\cite{Neumann} & Active & MAC & Type of device \\
\hline
Duration field values in 802.11 data and management frames\cite{Neumann} & Passive & MAC & Type of device \\
\hline
Clock skew calculated by using the timestamps contained in beacon frames\cite{Neumann} & Passive & MAC & Individual device \\
\hline
Traffic profiles & Passive & Application & Individual device \\
\hline
Clock skew calculated by using the message timestamps in PHY layer & Passive & PHY & Individual device \\
\hline
\end{tabularx}
\end{table}

\section{Fingerprinting Algorithms}
\label{sect:algo}

After extracting features and generating device fingerprints, the final and key step is to develop fingerprinting algorithms to identify wireless devices and detect illegitimate ones.  According to whether prior information of legitimate devices is required, fingerprinting algorithms can be divided into two categories: white-list based  and unsupervised learning based algorithms.  White-list based algorithms need to register legitimate devices to set up a database of the feature space of legitimate devices, while unsupervised learning based algorithms do not require such prior knowledge.

\subsection{White-list based algorithms}

Given a set of legitimate devices whose fingerprints (master signatures) are known, the device identification problem is thus to determine whether an unknown device is one of the legitimate devices based on its fingerprint.

\begin{figure}[ht!]
\centering
\includegraphics[width=0.8\textwidth]{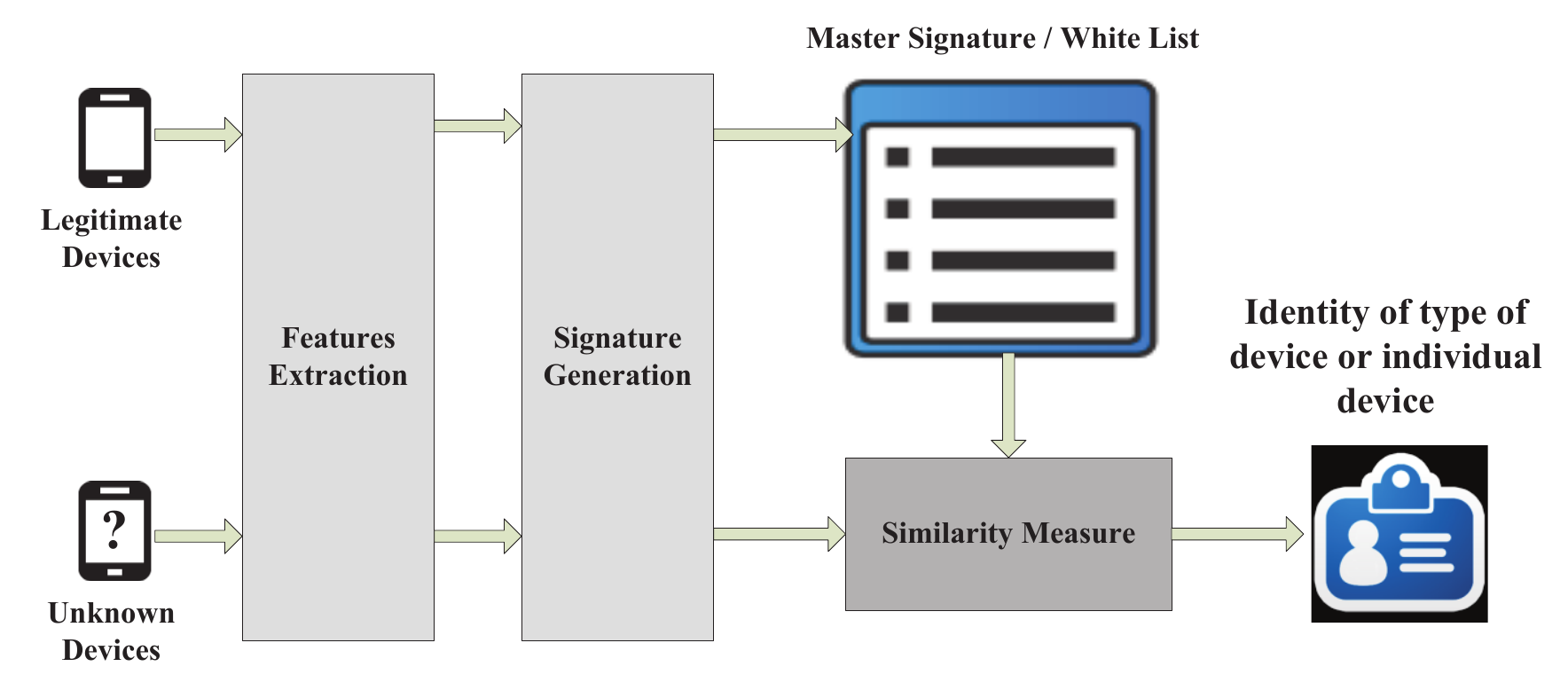}
\caption{White-list based fingerprinting algorithm using similarity measurement}
\label{supervised1}
\end{figure}

If the fingerprint of each device can be represented by a vector, one simple approach is to compute the similarity of the new fingerprint with each master signature and apply a threshold. One commonly used similarity metric for vectors is cosine similarity.  Given two vectors $a$ and $b$, their cosine similarity is defined as:
$$
\textrm{cos}(\theta) = \frac{a\cdot b}{\parallel a \parallel \cdot \parallel b \parallel},
$$
where $a\cdot b$ is the Euclidean dot product. Other similarity metrics can be adopted based on the characteristics of the features.  Figure~\ref{supervised1} illustrates the basic steps of such an approach.

Franklin {\it et al.}~\cite{Franklin} used the time deltas between probe requests characterized by a binning approach as feature. Similarity is computed by iterating through all bins, summing the difference of the percentages and mean differences scaled by the percentage. The identification accuracy is shown to vary from 77\% to 97\% depending on the bin size. Corbett \textit{et al.}~\cite{Corbett2008a} proposed a method using signal processing to analyze the periodicity embedded in wireless traffic caused by active scanning, and hence created a stable spectral profile of different types of devices. An evaluation was conducted and showed that the proposed method can distinguish between NICs manufactured by different vendors with zero false positives. In order to fingerprinting wireless device, Gao \textit{et al.}~\cite{Gao} proposed a passive blackbox-based technique to identify different type of devices. They extracted IAT as features and showed high accuracy under the experiment scenario which has 6 types of AP.

\begin{figure}[ht!]
\centering
\includegraphics[width=0.8\textwidth]{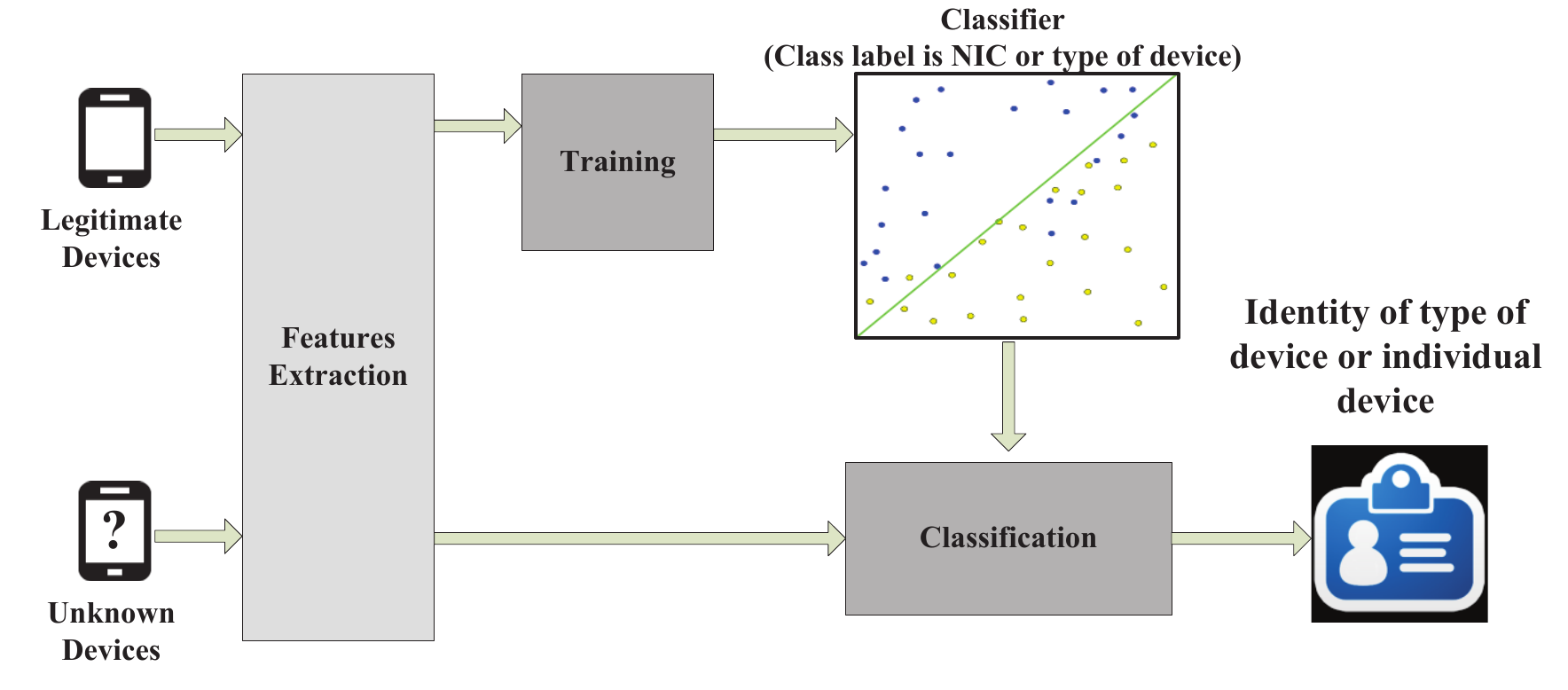}
\caption{White-list based fingerprinting algorithm using classification}
\label{supervised2}
\end{figure}

The master signature can sometimes be a set of fingerprints or can be represented using a distribution. In the case, we can consider each type (to identify different type of devices) or NIC ID (to identify individual devices) as a class, the problem is then converted into a typical classification problem: given a set of known classes, and a series of training data (fingerprints) with class labels, how to classify a new incoming device into these different classes? Traditional classification algorithms can be used, and the flow diagram is shown in Figure~\ref{supervised2}. There are several previous literatures adopt this approach as well~\cite{Bratus2008,Ureten2007,Brik2008,Scanlon2010,Hall2005}. By analyzing the time characteristics of 802.11 probe request frames, Desmond \textit{et al.}~\cite{Desmond2008} proposed a fingerprinting technique that differentiates between different individual devices. In their approach fingerprints of unique devices are represented by the clusters of inter-burst latencies which are processed timing measurements of probe requests during active scanning. To compare this kind of fingerprints the statistical hypothesis testing is employed to determine if different traffic traces captured are in fact emitted from the same or different devices.  In controlled environments, their technique was shown to be consistent and accurate in distinguishing among different devices. By challenging a user to prove its identity and hence reinforce the security of wireless network, Ureten and Serinken~\cite{Ureten2007} presented a complete identification system to identify an individual node in a wireless network. They exploited RF features complex amplitude and phase angle to generate fingerprints and used a probabilistic neural network for classification. Under the limited testing environment the designed system was able to classify signals with error rate $2\%$. As we discussed earlier, the PARADIS system designed and implemented by Brik \textit{et al.}~\cite{Brik2008} also adopted classification algorithm to identify network interface card of an IEEE 802.11 frame. They used radiometric features including frequency error, SYSNC correlation, I/Q offset, magnitude error and phase error, and classical classification algorithms including k-nearest neighbors (KNN) and support vector machine (SVM). PARADIS was shown to be able to differentiate among more than 130 NICs with an accuracy more than $99\%$.

\subsection{Fingerprinting using unsupervised learning}

\begin{figure}[ht!]
\centering
\includegraphics[width=0.9\textwidth]{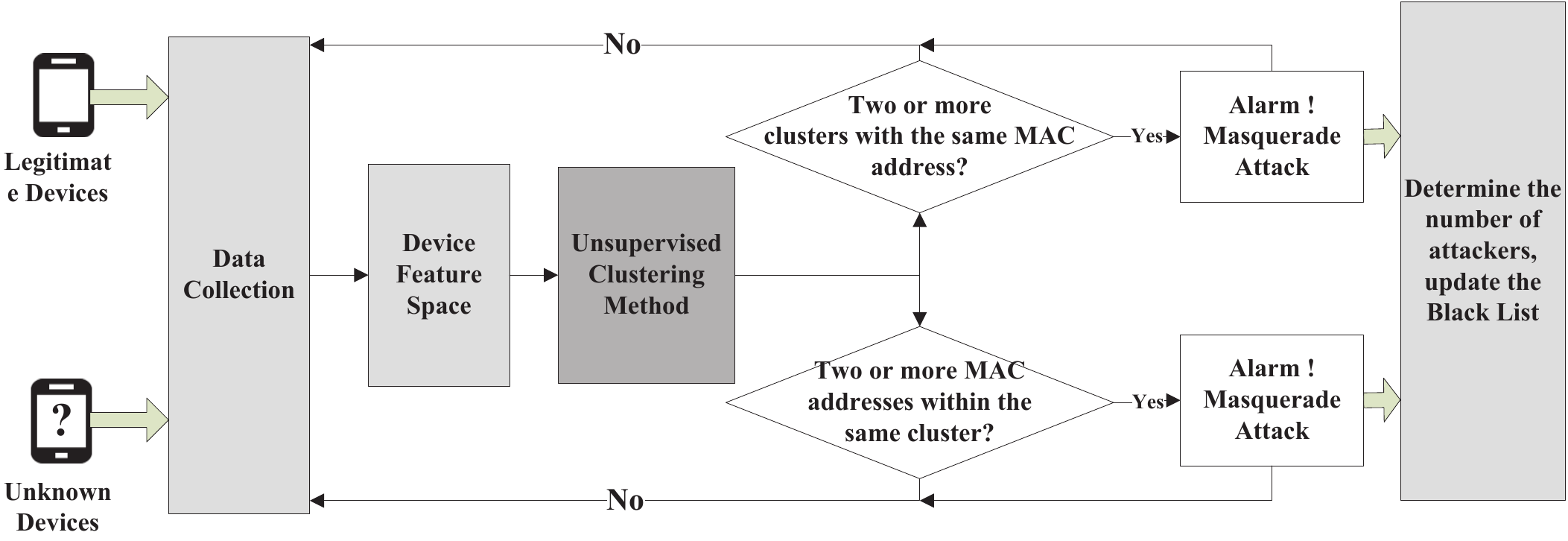}
\caption{The application of unsupervised learning fingerprinting algorithms to device fingerprinting}
\label{unsupervised}
\end{figure}

Even though white-list based fingerprinting algorithms are effective when the fingerprints of legitimate devices are available, registration of device fingerprints ahead of time is not always feasible in practice. Hardware upgrade, new purchasing or guest hosts introduce changes to the collection of legitimate devices making such approaches unscalable in enterprise environments. As an alternative, several recent works such as in~\cite{Nguyen2011,Chen} proposed the application of unsupervised learning in device fingerprinting. Unsupervised learning is a machine learning term used to describe the process to find hidden structures in unlabeled data. In the context of device fingerprinting, unsupervised learning based algorithms identify devices with similar fingerprints and cluster them together. Due to the lack of legitimate device information, unsupervised approaches generally cannot distinguish legitimate devices from illegitimate ones. However, they are effective in detecting the {\it presence} of identity spoofing when multiple devices with different fingerprints assume the same identifier (also known as {\it masquerade attack}), or a single device assumes multiple identifier (also known as {\it Sybil attacks}). The process of applying unsupervised learning algorithm to detect masquerade and Sybil attacks is shown in Figure~\ref{unsupervised}.
 
One key challenge in applying unsupervised learning algorithms to device fingerprinting is that the number of unique devices is not a given {\it prior}. Consequently, the number of clusters is unknown and needs to be extracted from the data. Nguyen {\it et al.}~\cite{Nguyen2011} is among the first that applied non-parametric Bayesian classification (NBC) approaches to device fingerprinting. NBC has the advantage of adapting the model complexity (i.e., the number of clusters) to the amount of data available, and thus avoiding the problem of over-fitting and under-fitting~\cite{bishop2006pattern}.

More concretely, in~\cite{Nguyen2011}, the feature space of a single device is modeled by a multivariate Gaussian distribution with unknown parameters. Depending on the the prior knowledge about the data, two classes of models can be used. The finite Gaussian mixture Model (FGMM) is used in the case where the number of clusters is known {\it a priori}, while the infinite Gaussian mixture model (IGMM) is used when the number of clusters is unknown or may vary over time. Each cluster is associated with a unique physical device, and thus the number of clusters equals to the number of active devices. Since in general, we have no control over the number of active (legitimate and illegitimate) devices, IGMM is more suitable. With IGMM, the fingerprint space of multiple physical devices (of the same or different device IDs) is modeled as an infinite Gaussian mixture (though only a finite subset can be observed). A non-parametric Bayesian approach to unsupervised clustering with an unbounded number of mixtures could be then developed. In the non-parametric Bayesian model, the cluster labels are generated using a Dirichlet process (DP) -- its realization being a conjugate prior of categorical distribution. Specifically, given the data set, $\vec{X}$, the hyper-parameters of the Gaussian mixtures, $H$ the base distribution of DP, and the concentration parameter $\alpha$, the goal is to determine the indicator $z_i$ of observation $i$. To do so, we need to derive an expression for the distribution of $\vec{Z}$ given the prior knowledge and then, use the Gibbs sampling method to sample from the distribution and find the class labels with the Maximum a Posterior.  The marginal distribution of each $z_i$ given by $P(z_i=k|\vec{Z}_{-i},\alpha,\vec{\theta},\vec{H},\vec{X})$, where $z_i$ is the unknown label of observation $i$ while $\vec{Z}_{-i}$ is the vector of labels to observations other than the $i$th one. With the application of the Bayesian rule, the marginal distribution can be written as:

\begin{equation}
\begin{array}{lll}
    P(z_i=k|\vec{Z}_{-i},\alpha,\vec{\theta},H,\vec{X}) & = & P(z_i=k|\vec{Z}_{-i},\alpha,\vec{\theta}_k,\vec{H},\vec{x}_i)  \\
    & \sim & P(\vec{x}_i|z_i = k, \vec{Z}_{-i}, \alpha, \vec{\theta}_k, H)P(z_i = k|\vec{Z}_{-i}, \alpha) \\
    & \sim & P(\vec{x}_i|\vec{\theta}_k)P(z_i = k|\vec{Z}_{-i}, \alpha),
\end{array}
\label{eq:nbc_posterior}
\end{equation}
where $\vec{X}$ is the set of feature points, $\vec{\theta}$ is the parameter of distribution of each cluster, $z_i = k$ indicates that the feature point $\vec{x_i}$ belongs to class $k$. In~(\ref{eq:nbc_posterior}), $P(\vec{x}_i|\vec{\theta}_k)$ is the likelihood and simply a Gaussian distribution. The only unknown term is $P(z_i = k|\vec{Z}_{-i}, \alpha)$, which can be determined by the generative model FGMM or IGMM. A Collapsed Gibbs Sampling method was proposed to reduce the sample space while sampling the values of the parameters and hence improve the efficiency of the algorithm.

In~\cite{Nguyen2011}, Frequency difference and phase shift difference have been used as device fingerprints. The authors conducted experiments using Zigbee devices in controlled experiments and achieved $98.2\%$ success rate in detecting Masquerade attacks and $99.3\%$ probability in successfully detecting Sybil attacks.

In addition to IGMM, the infinite hidden markov random field model (IHMRF) has also been applied in designing fingerprinting algorithms\cite{Chen}. Unlike \cite{Nguyen2011}, Chen {\it et al.}~\cite{Chen} use both location-independent features such as frequency difference and phase shift difference as well as location-dependent features such as RSS and angle of arrival (AoA). Given a set of observations {$(x_1,s_1), ..., (x_N, s_N)$}, where each observation $(x_i, s_i)$ has $p$ features $(x_i \in \mathcal{R}^p)$ and $d$ spatial coordinates $(s_i \in \mathcal{R}^d)$, the goal is to infer the latent variables $Z=\{z_i, ..., z_n\}$ based on $X$ and $S$, where $z_i \in \mathbb{C}$, and $\mathbb{C} = \{1,..., C\}$ denotes the set of class labels. In wireless fingerprinting, class labels correspond to the identities of wireless devices. The IHMRF model can be represented by a graphical model where the filled nodes refer to observations and blank nodes refer to latent variables $Z$. Given the time stamps $T=\{t_1, t_2, ...,t_N\}$, the spatial-temporal features \{$(s_1,t_1),...,(s_N,t_N)$\} were used to build a neighborhood graph for the latent states $Z$, in which states $z_i$ and $z_j$ are connected by an undirected edge if they are spatial temporal neighbors. Each latent state variable $z_i$ emits an observation $x_i$. The key difference between IHMRF and IGMM is that IHMRF utilizes the spatial-temporal vicinity of observations. Based on the hidden Markov random field theory, the hidden states should be consistent if they are neighbors to each other, while in the original IGMM model, two neighbor nodes $z_i$ and $z_j$ might follow different clusters if their emission observations $x_i$ and $x_j$ belong to different Gaussian distributions. The combination of these two independent estimates will improve the estimation accuracy. 

Compared with IGMM, IHMRF is able to capture the spatial dependencies between latent variables $\{Z_i\}^{N}_{i=1}$ concurrently by introducing $\beta$ (the inverse temperature of the model) and $\gamma$ (the inverse temperature parameter). The value of $z_i$ is decided based not only on its spatial neighbors but also on its closest Gaussian mixture. If an observation's neighbor's class label is not consistent with its closest Gaussian mixture, $\gamma$ is used to adjust the weight we put on each side. A smaller $\gamma$ implies a larger weight of the contribution of Gaussian mixture. In an extreme case that $\gamma = 0$, IHMRF degenerates to IGMM. For better scalability, variational method is used in~\cite{Chen} to implement the inference on IHMRF. Through simulation studies, the IHMRF model is shown to perform better than the IGMM model in precision, recall, F-measure and relative index.

\subsection{Summary}

In this section, we have reviewed existing algorithms on device fingerprinting.  Depending on whether prior knowledge of the fingerprints of legitimate devices is available, two types of approaches can be taken, namely, white-list based approaches and unsupervised learning based approaches. Compared to white-list based approaches, unsupervised learning based approaches incur higher computation complexity. Furthermore, they can only detect the presence of the attacks and the likely culprit while not being able to exactly pin down the malicious devices. Despite the limitations, unsupervised learning based approaches require no human intervention and pre-registration process, making them more practical in real deployments.

\section{Open Problems and Future Research Directions}
\label{sect:future}
In this section, we discuss some open problems and possible future research directions in wireless security using device fingerprints.

\subsection{Feature selection}

In Section~\ref{sect:feature}, we provided a taxonomy of features that can be used for device fingerprinting. As summarized in Table~\ref{big table}, features may be obtained from different layers of the network stack, and may work at different granularities. The important questions are, subject to hardware constraints, i) how to determine the best set of features to extract?  and ii) how to best combine the features for device fingerprinting? These are topics of feature selection.

Feature selection, also known as variable selection, attribute selection, is the process to select a subset of relevant features for use in model construction~\cite{Guyon2003}.  The key idea of feature selection is that there are many redundant or irrelevant features, and machine learning techniques such as principle component analysis (PCA) and SVM can be employed to reduce the dimension of the feature space, or project the features to a transformed space. There has been little work on feature selection for device fingerprinting. In \cite{Scanlon2010}, useful RF fingerprints were extracted from spectral components of the RACH preamble sequence across the transition phase as well as the steady-state portion of the signal. To reduce the dimensionality and remove irrelevant noisy features of the feature vector, an information theoretic approach to feature selection is explored. Mutual information is used as a basis for selecting a subset of all possible spectral features. Although only limited works have been done on fingerprints features selection, there are still several criteria can be adopted:

\begin{itemize}
\item PHY layer features may vary due to environmental factors or mobility of transceivers. It is thus crucial to select robust device-specific features that are less sensitive to device location and environmental settings. For example, features in modulation domain are typically more robust than those in waveform domain.

\item MAC layer features are limited to individual wireless technologies. While they are easier to extract using off-the-shelf components, MAC layer features are typically vendor specific and cannot be used for device fingerprinting. Another consideration of the choice of MAC or upper layer features is whether the associated feature is available from closed networks that employ data encryption (e.g., WPA, WPA2 in 802.11).

\item Features that can be extracted passively are generally preferred compared to active features.
\end{itemize}

\subsection{Benchmarking fingerprinting algorithms}

Many existing fingerprinting algorithms are evaluated in isolation through controlled experiments or using simulated or synthetic data. As a result, it is hard to compare their performance objectively. Publicly available data repository such as CRAWDAD~\cite{Crawdad}  can help alleviate such a problem but the datasets do not contain traces from illegitimate devices. Furthermore, most datasets are either MAC layer traces or ill-suited for extracting rich PHY layer features.

To benchmark fingerprinting algorithms, researchers need to take a multi-pronged approach. First, as a community, we need a comprehensive collection of datasets for evaluating different algorithms. The datasets should contain labeled traces from a variety of devices collected from both indoor and outdoor environments, and different mobility patterns. Both raw signals and extracted features shall be included. Second, in addition to conventional performance metrics such as confusion matrix, Receiver operating characteristic (ROC), precision and recall rate, the scalability and execution time of the proposed algorithms  need to be assessed as well. Algorithms that are computational intensive may not be suitable in scenarios where a large number devices may be present and timely decisions are needed. Another relevant metric is energy consumption, which is crucial in portable systems. Third, the attacker model should not be limited to identity spoofing. One interesting problem is whether the fingerprinting algorithms themselves are resilient to attacks. For instance, if it is known that traffic patterns in WLANs are used as fingerprints, an attacker may selectively jam the channel to modify the packet inter-arrival time and deceive the fingerprinting algorithm into misidentification.

\subsection{Fingerprinting non-WiFi devices}

Predominantly, existing device fingerprinting approaches target WLAN networks. Most known MAC layer features pertain to 802.11 frames. With the pervasive deployment of wireless technologies in wide-area data networks, point-of-sale systems, and localization systems, just to name a few, it is expected that attacks on non-WiFi wireless devices and networks such as bluetooth low-energy (BLE), cellular systems, near-field communication (NFC) devices may pose serious threats with severe financial and privacy ramifications. NFC is a special type, but Ultra-high-frequency (UHF) radio frequency identification (RFID) fingerprinting is important, because for RFID authentication and cryptography are challenging.

\subsection{Other topics}
There are a few other research directions that are worth exploring. First, existing device fingerprinting approaches are primarily concerned with improving the accuracy of detection. This may be at the expense of high computation complexity and higher energy costs at the data collectors. Second, it will be interesting to design algorithms that require as few observations as possible to limit the negative impact of malicious users. This is in analogous to quickest detection approaches for anomalies or changes \cite{poor2009quickest}. However, the problem is more complex since the observations are multidimensional and the fingerprints of illegitimate users are unknown a {\it priori}. Third, while wireless fingerprints are important in device identification, other sources of signals may be utilized as well including acoustic, thermal or magnetic signatures. Considering the rapid innovation pace in wireless fingerprinting scenario, the area of device fingerprinting will continuously face numerous challenges that cannot be covered within the scope of this paper.

\section{Conclusions} 
\label{sect:conclusion}
In this paper, we have provided a comprehensive survey on device fingerprinting for enhancing wireless network security. A taxonomy of features that can be used for fingerprinting, along with several fingerprinting algorithms have been elaborated.The key idea is to extract characteristics from transmitted signal or frames from the wireless devices and their environments to generate non-forgeable signatures. The unique signatures are then used to distinguish between legitimate and malicious devices. Device fingerprinting solutions can be combined with other techniques such as localization and tracking to mitigate insider attacks in wireless networks. 

\bibliographystyle{IEEEtran} 
\bibliography{MyFingerprint} 
\end{document}